\documentclass[%
reprint,
nofootinbib,
 amsmath,amssymb,
 aps,
]{revtex4-1}
\usepackage{graphicx}
\usepackage{dcolumn}
\usepackage{bm}
\usepackage{hyperref}

\usepackage{subcaption}
\usepackage[font=small,labelfont=bf,
   justification=Justified,
   format=plain]{caption} 
\usepackage{amsmath}
\usepackage{bbm}
\usepackage{amsfonts}
\usepackage{amssymb}
\usepackage{latexsym}
\usepackage{graphicx}
\usepackage[english]{babel}
\usepackage{multirow}
\usepackage{float}
\usepackage{url}
\usepackage{slashed}
\usepackage{xcolor} 
\usepackage[utf8]{inputenc}
\usepackage{verbatim}
\usepackage{stackengine}
\usepackage{comment}

\usepackage{cleveref} 
    \crefname{equation}{}{}
    \crefname{figure}{}{}    
    \crefname{table}{}{}
    \crefname{section}{}{}   
    \crefname{appendix}{}{}
    \crefname{footnote}{}{}
    
    \crefalias{subequation}{equation}

\newcommand{\be}{\begin{equation}}
\newcommand{\ee}{\end{equation}}
\newcommand{\ba}{\begin{array}}
\newcommand{\ea}{\end{array}}
\newcommand{\bea}{\begin{eqnarray}}
\newcommand{\eea}{\end{eqnarray}}

\newcommand{\nn}{\nonumber}

\usepackage{slashed}

\begin{document}


\title{Bubble wall velocity for first-order QCD phase transition}

\author{James M. Cline}
\email{jcline@physics.mcgill.ca}
\thanks{ORCID: \href{https://orcid.org/0000-0001-7437-4193}{0000-0001-7437-4193}}
 \author{Benoit Laurent}
\email{benoit.laurent@mail.mcgill.ca}\thanks{ORCID: \href{https://orcid.org/0000-0002-1306-3620}{0000-0002-1306-3620}}
\affiliation{McGill University Department of Physics \& Trottier Space Institute, 3600 Rue University, Montr\'eal, QC, H3A 2T8, Canada}
 \affiliation{McGill University, Department of Physics}

\begin{abstract}
Although the QCD phase transition is a crossover in the standard model, nonstandard effects such as a large lepton asymmetry are known to make it first order, with possible applications to gravitational wave production.  This process is sensitive to the speed of the bubble walls 
during the phase transition, which is difficult to compute from first principles. 
We take advantage of recent progress on wall speed determinations to provide a simple estimate valid in the small supercooling regime which constrains the wall speed to be significantly lower than what has been used in previous literature.  This in turn strongly suppresses the production of gravitational waves, to a level that is just out of reach of the most sensitive projected experiment for 
this signal, $\mu$Ares. While our analysis approximates the equation of state using the template model, we demonstrate that our conclusions remain robust when incorporating state-of-the-art QCD equation of state data.

\end{abstract}
\maketitle

Cosmological phase transitions are an active subject of study, in particular those which are first order and thus proceed by the nucleation of true-vacuum bubbles in a supercooled false-vacuum plasma.  Provided the supercooling is not
too extreme, the bubbles will expand, reach some terminal velocity, and collide with each other until the Universe is entirely in the new true vacuum state.
During the epoch of bubble expansion, the baryon asymmetry of the universe could be created, if sphaleron interactions are sufficiently fast and there is a source of CP violation.  Gravitational waves are produced by the bubble wall collisions and sound waves propagating in the plasma.

The efficiency of both processes depends upon the terminal speed $v_w$ of the bubble walls, which could be as large as $1$ (the speed of light) if the friction of the plasma on the wall is small compared to the pressure difference across the wall.  However in a confinement/deconfinement transition as for QCD, there is typically a large change in the number of light degrees of freedom in the plasma when crossing the wall, which can lead to substantial friction and a relatively low wall speed. (For QCD, there are $\sim$51 light degrees of freedom above the critical temperature, and $\sim$17 below.)
To determine $v_w$ precisely is difficult, since it requires
solving integro-differential equations governing the wall dynamics and deviations of particle distributions from their thermal values in the vicinity of the wall
\cite{Laurent:2022jrs,Ekstedt:2024fyq,Dorsch:2024jjl}.

Recently, methods for estimating bubble wall speeds using simpler techniques have been developed \cite{Ai:2021kak,Ai:2023see,Sanchez-Garitaonandia:2023zqz,Ai:2024btx,Krajewski:2024zxg}.  These methods identify extreme cases to put upper and lower bounds on $v_w$.  In some cases, the two bounds are close enough to each other that a relatively quantitative estimate for the actual value of $v_w$ is obtained without having to undertake formidable coding tasks or numerical computations.
Our goal in the present note is to carry this out for the QCD phase transition,
in nonstandard scenarios such as large lepton asymmetries \cite{Schwarz:2009ii,Wygas:2018otj,Gao:2021nwz,Gao:2024fhm,Gao:2023djs} that can render it first order in the early Universe.\footnote{Another possibility is to make the QCD coupling dynamical so that
the QCD transition occurs at temperatures above the electroweak scale
\cite{Ipek:2018lhm}.  In this case the quarks are still massless, which leads to a first-order QCD phase transition.  However it occurs at the TeV scale, and a separate analysis from the present one would be needed to determine the wall velocity in this scenario.}  Even though the baryon chemical potential is small in the present universe, during the QCD epoch, a large lepton asymmetry temporarily
induces large quark chemical potentials through the weak interactions, which can change the order of the 
QCD phase transition.
The bubble wall velocity was also estimated for a general $SU(N)$ confinement phase transition in \cite{Gouttenoire:2023roe}.

{Lattice simulations have shown that for vanishing chemical potential, in the (2+1)-flavor case (which is relevant for transitions happening at $T\sim 100\ \mathrm{MeV}$), the phase transition is a crossover
\cite{Wu:2006su,JLQCD:1998mja,Guenther:2020jwe}. However, when a sufficiently large baryon chemical potential $\mu_B$ is present, it may become first order. A combination of lattice and functional QCD studies have shown that the transition line can be parametrized by
\be
\frac{T_c(\mu_B)}{T_c(0)}=1-\kappa\left(\frac{\mu_B}{T_c(0)}\right)^2+\cdots
\ee
with $T_c(0)=155\ \mathrm{MeV}$ and $\kappa=0.016$ \cite{Lu:2023msn}. Furthermore, a critical end point (CEP) is expected to be located at $(T_c^{\rm CEP},\mu_B^{\rm CEP})\approx(118, 600)\ \mathrm{MeV}$, with the phase transition (PT) becoming first order when $\mu_B>\mu_B^{\rm CEP}$. In the scenario considered here, we assume the existence of a large transient baryon asymmetry (as has been shown to arise in the presence of large lepton asymmetries \cite{Schwarz:2009ii}) to satisfy this condition, causing the QCD PT to be first order and proceeding via bubble nucleation.
}

A number of previous works have considered the production of gravitational waves from a first-order
QCD phase transition.  Most of them assume that the 
bubbles expand with speeds close to 1
\cite{Shao:2024dxt,Zheng:2024tib,Han:2023znh,Gao:2023djs,
Feng:2022fwf,Zhu:2021vkj,Rezapour:2020mvi,Caprini:2010xv,Ahmadvand:2017xrw,Reichert:2021cvs}, or possibly as low as $v_w=0.2$ \cite{Huang:2020crf,Shao:2022oqw}.  Those references that allowed for lower values of $v_w$, such as would occur for {weak} deflagrations, found that no observable gravitational waves (GWs) would be produced \cite{Li:2018oqf,Ahmadvand:2017tue}. Using the template model, which assumes a uniform speed of sound, to approximate the full QCD equation of state, we will argue that the dynamics of the QCD phase transition are such that we are always in this situation, where GW production is suppressed to unobservable levels, at least for the currently envisioned future experiments.

\section{Bubble nucleation}

Computing the bubble wall velocity of a first-order phase transition (FOPT) requires knowing the value of the pressure difference $\Delta p$ between the two sides of the wall during the phase transition. Effectively, it was shown in Ref.\ \cite{Ai:2023see} that for weak FOPTs (which lead to slowly expanding bubbles), the wall velocity in local thermal equilibrium scales like $v_w\propto\sqrt{\Delta p}$. 
To compute it, one must evaluate the bubble nucleation rate per volume $\Gamma$. The percolation temperature $T_*$ and the pressure difference can then be obtained by requiring that most of the Universe is in the new phase.

As there is no scalar potential for the QCD phase transition, it may be challenging to compute the tunneling rate with the standard theory of false-vacuum decay which consists in finding the bounce action. However, similar to Ref.\ \cite{Gao:2024fhm}, we can use classical nucleation theory to obtain a simple estimate.

The energy of a single bubble of size $R$ is given by
\be
E(R) = -\frac{4\pi}{3}R^3\Delta p +4\pi R^2\sigma,
\ee
with $\sigma$ the surface tension.
As the system wants to minimize its energy, a bubble can only expand if $dE/dR<0$. This requires the bubble to be larger than the critical radius at which $dE/dR = 0$,
\be
R_c=\frac{2\sigma}{\Delta p}\,.
\ee
The energy of a critical bubble is then
\be\label{eq:criticalEnergy}
E(R_c)=\frac{16\pi\sigma^3}{3(\Delta p)^2}\,.
\ee
Assuming that the plasma configuration is sampled from a Boltzmann distribution, the nucleation rate per volume of a critical bubble is then given by
\be\label{eq:nucleationRate}
\Gamma(T)=AT^4e^{-E(R_c)/T},
\ee
with $A$ some prefactor that we will take to be of order 1.

By integrating the nucleation rate, one can show that most of the Universe is in the true vacuum phase when \cite{Huber:2007vva} 
\bea
\frac{E(R_c)}{T_*} &=& \log\left(\frac{8\pi A v_w^3 T_*^4}{\beta^4}\right) \\
&\approx& 117-4\log\left(\frac{T_*}{118\ \mathrm{MeV}}\right)-4\log\left(\frac{\beta/H}{10^5}\right)\nn\\
&&+\log(A)+3\log\left(\frac{v_w}{0.01}\right)-2\log\left(\frac{g_*}{51}\right),\nn
\eea
with $\beta$ the inverse PT duration which will be defined below, $H$ is the Hubble rate, and $g_*$ the effective number of degrees of freedom. In what follows, we will neglect the weak dependence on the logarithms and use the simple estimate
\be\label{eq:nucleationCondition}
\frac{E(R_c)}{T_*}\approx 117\,,
\ee
which is valid at the QCD scale.  Combining Eqs.\ (\ref{eq:criticalEnergy}) and (\ref{eq:nucleationCondition}), one obtains the pressure difference at the percolation temperature $T_*$,
\be\label{eq:pressureDiff}
\Delta p(T_*)\approx 0.38\sqrt{\sigma^3(T_*)/T_*}\,.
\ee

The inverse duration of the PT,
\be
\beta = \frac{d\log\Gamma}{dt}\approx HT\frac{d(E/T)}{dT}\,,
\ee
is important  for determining the spectrum of GWs that can be produced by collisions of the bubbles.
For PTs with a small amount of supercooling, $\Delta p(T)$ can be linearized around the critical temperature $T_c$,
\be\label{eq:eosApprox}
\Delta p(T)\approx a(T_c)(T_c-T)\,,
\ee
where $a(T_c)$ is a function that will be determined below from the equation of state
(EOS)  and $T_c$ is defined by $\Delta p(T_c)=0$.  The critical bubble energy $E(R_c)$ therefore diverges as $(T_c-T)^{-2}$, and this dependence gives the dominant  contribution to $d(E/T)/dT$
near $T_c$. Hence 
\bea
\frac{\beta(T_*)}{H(T_*)}&\approx& \frac{2E(T_*)}{T_c-T_*}=\frac{32\pi\, a(T_c)\,\sigma^3(T_*)}{3(\Delta p(T_*))^3}\nn \\
&\approx& 618 a(T_c)\left[T_*/\sigma(T_*)\right]^{3/2}.
\eea

This can be simplified further by assuming that the plasma can be approximated by the template model \cite{Leitao:2014pda}, which is a generalization of the bag EOS, allowing the plasma to have an arbitrary (but $T$-independent) sound speed $c_s$. The pressure, energy and enthalpy in this EOS are
\bea\label{eq:templateModel}
p(T) &=& \frac{b}{3}T^\nu T_*^{4-\nu}-\epsilon,\\
e(T) &=& \frac{b}{3}(\nu-1)T^\nu T_*^{4-\nu}+\epsilon,\\
w(T) &=& \frac{b}{3}\nu T^\nu T_*^{4-\nu}\,,
\eea
where $\nu=1+1/c_s^2$, $\epsilon$ is the vacuum energy and $b=\pi^2 g_*/30$. Although it is generally possible to have a different sound speed on both sides of the wall, we will assume here that it is the same to simplify the analysis. This model reduces to the standard bag EOS when $\nu=4$.

Within the template model, one can show that $a(T_c)=\nu\pi^2\Delta g_* T_c^3/90$, which implies 
\be
\frac{\beta(T_*)}{H(T_*)}\approx 67.8\frac{\nu\Delta g_* T_c^{9/2}}{\sigma^{3/2}(T_c)},
\label{betaH}
\ee
where we have approximated\footnote{From Eqs.\ (\ref{eq:pressureDiff}) and (\ref{eq:eosApprox}), it is straightforward to show that, to lowest order in $T_c-T_*$, $T_*\approx T_c-\frac{0.38}{a(T_c)}\sqrt{\sigma^3(T_c)/T_c}\approx T_c[1-0.025(\sigma(T_c)/T_c^3)^{3/2}]$. Therefore, the approximation $T_*\approx T_c$ is expected to be valid as long as $\sigma(T_c)/T_c^3\lesssim1$, which is satisfied for $T_c\gtrsim 84\ \mathrm{MeV}$.} $T_*\approx T_c$ (which is valid for a low degree of  supercooling) and $\Delta g_*\sim 34$ is the difference of effective degrees of freedom between the two phases. We will use $c_s^2=1/3$ (or $\nu=4$) to compute the numerical estimates throughout the rest of this work. 

Instead of using the template EOS, it is also possible to use a state-of-the-art QCD EOS such as those that have been determined in Refs.\ \cite{PhysRevC.100.024907,Lu:2023msn,Bresciani:2025vxw}.  For simplicity, we present the analytic results coming from the simpler template EOS in the main text.  In 
Appendix \ref{app:qcdeos}, we demonstrate that very similar results are found when using a more accurate
QCD EOS.

\section{Wall velocity}

It is challenging to precisely determine the wall velocity of a QCD FOPT, since it depends on the EOS and the deviation from equilibrium, which are difficult to compute. However, one can relatively easily bracket $v_w$ between robust lower and upper limits that apply in different regimes of friction on the wall.
It was shown in Ref.\ \cite{Ai:2024btx} that an upper bound on $v_w$ follows from making the local thermal equilibrium (LTE) approximation, and a lower bound is obtained in the ballistic limit. In this section, we will use these two approximations to derive bounds on $v_w$ for a plasma described by the bag EOS.

Within the bag EOS, both limits can be described in terms of two simple parameters. The strength of the phase transition is characterized by \footnote{This definition of $\alpha$ differs from the most common one found in the literature, which is defined in terms of the trace of the energy-momentum tensor $\theta=e-3p$ as $\alpha_\theta=\Delta\theta/(3w_+)=\alpha_p+(1-\psi)/3$. Although the equations can be written in terms in either definition equivalently, we find that $\alpha_p$ is more natural for describing the wall velocity of weak PTs, since $v_w$ goes to zero in the limit $\alpha_p\to0$.}
\be\label{eq:alphaP}
\alpha_p=\frac{4\Delta p}{3w_+},
\ee
where $w_+=\nu\pi^2g_{*}^+T_*^4/90$ is the enthalpy in front of the wall (in the template model). Using Eq.\ (\ref{eq:pressureDiff}), $\nu\approx 4$ and $g_*^+\sim51$, this reduces to
\be
\alpha_p\approx 0.023\left(\frac{\sigma}{T^3}\right)^{3/2}\,.
\ee
One also needs the ratio of enthalpies, which approximately corresponds to the ratio of effective degrees of freedom on both sides of the wall: 
\be\label{eq:psi}
\psi=\frac{w_-}{w_+}\approx \frac{g_{*}^-}{g_{*}^+}\approx 0.33.
\ee

Two estimates of the wall velocity were computed in Appendix \ref{app:hydro} using the LTE \cite{Ai:2023see} and ballistic \cite{Ai:2024btx} approximations. For weak PTs, the LTE limit, which gives an upper bound on the wall velocity is
\be
v_w^{\rm LTE}\approx \frac{c_s}{1-\psi}\sqrt{\frac{2\alpha_p(T_*)}{1+c_s^2}}\approx 0.16\left(\frac{\sigma}{T^3}\right)^{3/4}.
\ee
On the other hand, a lower bound can be obtained from the ballistic approximation which assumes that all the degrees of freedom that get confined during the PT cannot enter the bubble (which would be the case if the bound states are much heavier than $T$) and are thus reflected in front of the bubble. For weak PTs, this limit yields the wall velocity
\be
v_w^{\rm ball}\approx \frac{\alpha_p}{\psi(1-\psi)}\approx 0.1\left(\frac{\sigma}{T^3}\right)^{3/2}.
\label{vwballeq}
\ee

\begin{figure}
    \centering
    \includegraphics[width=0.9\linewidth]{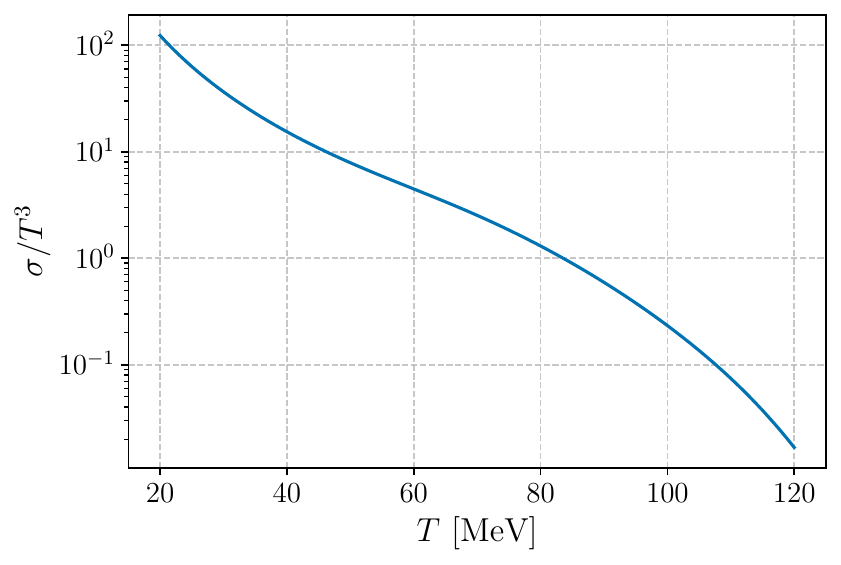}
    \caption{Fit of the QCD surface tension $\sigma/T^3$ computed in Ref.\ \cite{Gao:2016hks}.}
    \label{fig:sigma}
\end{figure}

We note that both estimates only depend on the ratio $\sigma/T^3$ and are valid when it is small. The QCD surface tension was evaluated in Ref.\ \cite{Gao:2016hks} using a Dyson-Schwinger equation approach. The numerical fit of this function is shown in Fig.\ \ref{fig:sigma} and is given by
\be
\sigma(T) = a+b\exp(c/T+d/T^2),
\ee
with
\bea
a&=& 9.89\times 10^5\ \mathrm{MeV}^3,\qquad b=-5.84\times 10^4\ \mathrm{MeV}^3\nn\\
c&=& 736\ \mathrm{MeV},\qquad d=-4.8\times 10^4\ \mathrm{MeV}^2\nn.
\eea
With these values, $\sigma/T^3< 1$ for temperatures $T\gtrsim 84\,$MeV.

\section{GW spectrum}
The dominant source of GWs produced by weak FOPT is sound waves. Simulations have shown that this produces a spectrum with a model-independent shape, having a peak amplitude and frequency that depend on just a few parameters, namely $v_w$, $T$, $\beta/H$, $g_*$ and $K$, the kinetic energy fraction (defined below).

For the PTs considered here, one can show the shock formation timescale is always much smaller than a Hubble time. In that case, simulations find that the GW spectrum is well approximated by the formula \cite{Guo:2020grp,Hindmarsh:2020hop}
\be\label{eq:GWSpectrum}
h^2\Omega_{\rm GW}(f)=4.3\times10^{-6}\!\left(\frac{51}{g_*}\right)^{\!\!1/3}\!\!\!\!\!K^{3/2}\!\left(\frac{H}{\beta}\right)^{\!\!2} \!\!v_w^2 C\!\!\left(\frac{f}{f_{\rm p,0}}\right),
\ee
where $C$ is a function that parametrizes the spectrum's shape and $f_{\rm p,0}$ is the peak frequency. They are given by
\bea
C(s)&=&s^3\left(\frac{7}{4+3s^2}\right)^{7/2}\,,\\
f_{\rm p,0} &=& \frac{3.3\times10^{-9}\ \mathrm{Hz}}{v_w}\left(\frac{\beta}{H}\right)\!\!\left(\frac{T}{118\ \mathrm{MeV}}\right)\left(\frac{g_*}
{51}\right)^{\!\!1/6}\!\!\!\!\!\!.
\eea

The only remaining parameter to evaluate for a complete GW spectrum prediction is $K$. It is defined as the ratio of kinetic energy density produced by the PT to the total energy density before the PT. For weak PTs with a slow-moving wall, we
show in Appendix \ref{app:hydro} that one can approximate $K$ by 
\be
K=\frac{3\nu(1-\psi)^2 v_w^2}{\nu-\psi}\approx 1.5\,v_w^2,
\ee
which gives a peak GW amplitude proportional to $v_w^5$:
\be\label{eq:GWAmplitude}
h^2\Omega_{\rm GW}(f_{\rm p,0})\sim 7.6\times10^{-6}\left(\frac{51}{g_*}\right)^{1/3}\left(\frac{H}{\beta}\right)^2 v_w^5\,.
\ee

We can now obtain an upper and lower bound on the GW amplitude by inserting the LTE and ballistic estimates of the wall velocity into Eq.\ (\ref{eq:GWAmplitude}) and using Eq.\ ({\ref{betaH}):
\bea
\label{eq:maxGW}
h^2\Omega_{\rm GW}^{\rm LTE}(f_{\rm p,0})&\sim& 9.3\times10^{-18}\left(\frac{\sigma}{T^3}\right)^{27/4},\\
\label{eq:minGW}
h^2\Omega_{\rm GW}^{\rm ball}(f_{\rm p,0})&\sim& 9.8\times10^{-19}\left(\frac{\sigma}{T^3}\right)^{21/2}
\eea
Notice that these estimates are only valid for weak PTs, which correspond to $\sigma/T^3\lesssim 1$. This bound is satisfied for temperatures close to the CEP temperature $T_{\rm CEP}=118\ \mathrm{MeV}$ with $\sigma_{\rm CEP}/T_{\rm CEP}^3\approx0.023$ and it remains valid down to $T\approx 83.6\ \mathrm{MeV}$.  The CEP is the point in the plane of temperature versus quark chemical potential at which the QCD phase transition goes from being first order to a crossover.

We now see how important it is to have an accurate estimate of the wall velocity. As a comparison, many references (e.g.\ \cite{Gao:2024fhm,Zheng:2024tib,Han:2023znh,Gao:2023djs,
Feng:2022fwf,Zhu:2021vkj,Rezapour:2020mvi,Caprini:2010xv,Ahmadvand:2017xrw,Shao:2022oqw}) simply assumed $v_w=0.3$ or $v_w=1$ which yields the kinetic energy fraction $K\sim 0.04$. Since the GW amplitude is proportional to $K^{3/2}v_w^2$, this simple assumption therefore leads to overestimating it by at least $0.04^{3/2}/((1.5v_w)^{3/2}v_w^2)$, which can become as large as $5\times10^7$ (LTE) or $10^{15}$ (ballistic) at the CEP. Even at the lowest temperature considered here ($T=84\ \mathrm{MeV}$), this simple assumption would lead to an overestimation of the GW amplitude by a factor of $44$ (LTE) or $985$ (ballistic).

Similar to Eqs.\ (\ref{eq:maxGW}) and (\ref{eq:minGW}), one can estimate the maximal and minimal values of the peak frequency as
\bea
f_{\rm p,0}^{\rm max} &\sim& 1.9\times10^{-4}\ \mathrm{Hz}\times\left(\frac{T^3}{\sigma}\right)^3\left(\frac{T}{118\ \mathrm{MeV}}\right),\\
f_{\rm p,0}^{\rm min} &\sim& 3\times10^{-4}\ \mathrm{Hz}\times\left(\frac{T^3}{\sigma}\right)^{9/4}\left(\frac{T}{118\ \mathrm{MeV}}\right)
\eea

\begin{figure}
    \centering
    \includegraphics[width=1\linewidth]{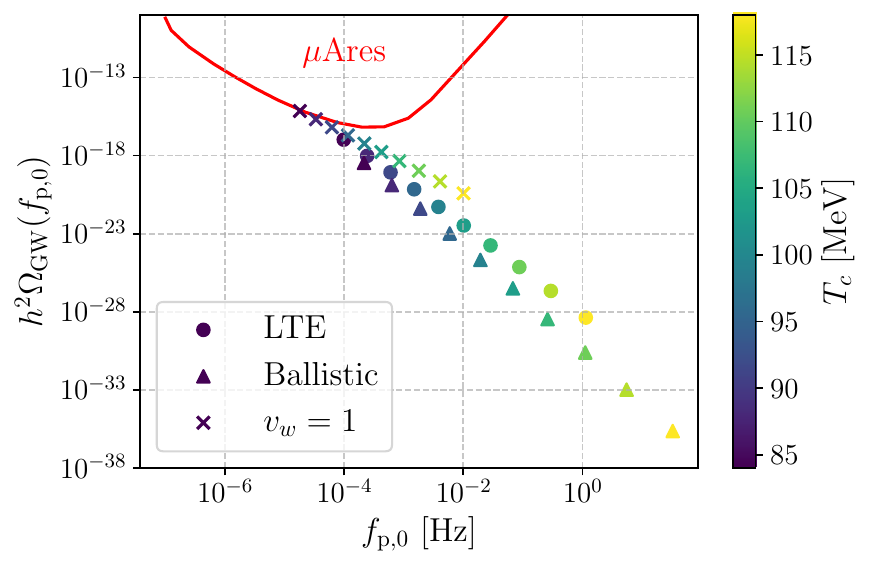}
    \caption{Position of the GW spectrum's peak for critical temperature in the range $T_c\in [84,118]\ \mathrm{MeV}$. We show the predictions obtained by estimating the wall velocity with the LTE and ballistic approximations and by simply assuming $v_w=1$. The red line shows the $\mu$Ares power law sensitivity curve.}
    \label{fig:GWAmplitude}
\end{figure}

We show in Fig.\ \ref{fig:GWAmplitude} the position of the GW spectrum's peak for wall velocities estimated using the LTE and ballistic approximation and for the simple assumption $v_w=1$. We explore the range of critical temperatures $T_c\in[84,118]\ \mathrm{MeV}$. Stronger PTs could be obtained at lower temperatures, but this would lead to more supercooling characterized by $\sigma(T_c)/T_c^3 > 1$, which would make the classical nucleation theory used to derive the nucleation rate (\ref{eq:criticalEnergy}) and (\ref{eq:nucleationRate}) unreliable. We compare the GW amplitude to the sensitivity of the proposed detector $\mu$Ares \cite{Sesana:2019vho}, which is meant to cover the gap between LISA and the pulsar timing arrays in the $\mu$Hz frequency band.

One can see that, in the temperature range considered, it is never a good approximation to assume $v_w=1$. At best, it overestimates the GW amplitude by two orders of magnitude while underestimating the peak frequency by a factor of 5 compared to the more realistic LTE or ballistic assumptions. And as expected, the discrepancy becomes far worse as $T_c$ gets closer to the CEP, since the weaker PT leads to slower walls.

With a realistic estimation of the wall speed, we found no models that could be seen by $\mu$Ares. There might still be a possibility of a QCD PT being probed if it happens at a lower temperature than what was considered here, which would require a larger quark chemical potential ($\mu_q \gtrsim 275\ \mathrm{MeV}$). Effectively, a lower critical temperature would produce a stronger PT and GW signal. However, as we previously mentioned, the classical nucleation theory would not be appropriate for describing these stronger PTs. Exploring cooler QCD PTs would therefore require a more detailed analysis, which we leave for future work.

\section{Summary and conclusions}
\label{sec:conc}

In this work we revisited the production of GWs during a QCD confinement phase transition, when it is made first order by new physics effects that could temporarily create sizable quark chemical potentials. Since the effective number of degrees of freedom changes significantly when quarks and gluons become confined, we expect the friction of the plasma on the bubble wall to be enhanced, leading to a suppressed rate of expansion of the bubble walls during the phase transition.

We have shown that, for slow-moving walls, the amplitude of the GWs scales as $v_w^5$. Hence, it is critical to reliably estimate of the bubble wall velocity for quantifying the GW spectrum produced during the phase transition. Although it is challenging to  precisely compute the wall velocity, due to the strongly-coupled nature of the underlying physics, simplifying approximations exist that yield rigorous bounds on  $v_w$.
 In particular, we have shown that the local thermal equilibrium  limit gives an upper bound going as $\sqrt{\Delta p}$, while the ballistic result (lower bound) is proportional to $\Delta p$, where $\Delta p$ is the pressure difference between the two sides of the wall. Note that the latter result is consistent with previous studies of strongly-coupled system which used holography to compute the bubble wall velocity \cite{Bea:2021zsu,Bigazzi:2021ucw}.

This underlines the need to quantify the degree of supercooling, which determines $\Delta p$. We did so  by estimating the nucleation rate using classical nucleation theory, which predicts the percolation temperature $T_*$, $\Delta p(T_*)$, and other important parameters such as $\beta/H$ (the duration of the phase transition). This method is expected to be valid in the range of critical temperatures $T_c\in[84,118]\ \mathrm{MeV}$, which implies a low degree of supercooling.

Our results show that over-optimistically 
assuming the wall velocity to be of $\mathcal{O}(1)$ overestimates the GW amplitude by factors as large as $10^{15}$, rendering such predictions completely unreliable. With our more realistic estimates of $v_w$, we found that the QCD phase transition is out of reach of the next-generation GW detector $\mu$Ares in the temperature range considered here. Stronger GW signal could be reached at lower values of $T_c$, but this would require a more careful analysis of the nucleation rate, which we leave for future work.

For simplicity, we used the bag equation of state to derive the preceding results. One might naturally question whether they could be altered by using
a realistic QCD EOS, such as has been numerically determined in the literature. We have checked that this is not the case, and that our conclusions remain the same when using state-of-the-art EOS data from Ref.\ \cite{Lu:2023msn}.

\bigskip
\section*{Acknowledgments.} We thank Fei Gao for his assistance in obtaining the QCD EOS data. JC is  supported by NSERC
(Natural Sciences and Engineering Research Council, Canada). BL is supported by the Fonds de recherche du Québec Nature et technologies (FRQNT).
\begin{appendix}

\section{Hydrodynamics at small $v_w$}
\label{app:hydro}

We describe here how the fluid around the wall can be described in the small-$v_w$ limit. We will assume that the plasma's equation of state can be described by the template model defined in Eq.\ (\ref{eq:templateModel}).
The fluid velocity and temperature profiles can be obtained by requiring conservation of energy and momentum. Across the wall, this implies \cite{Giese:2020rtr}
\be\label{eq:matching}
\frac{v_+}{v_-} = \frac{v_+ v_-/c_s^2-1+3\alpha_\theta(T_+)}{v_+ v_-/c_s^2-1+3v_+ v_-\alpha_\theta(T_+)}\,,
\ee
where the subscripts $+$ and $-$ refer to the quantity in front and behind the wall, respectively. The fluid velocities $v_\pm$ are measured in the wall frame and
\bea
\alpha_\theta(T) &=& \frac{\theta_+(T)-\theta_-(T)}{3w_+(T)} = \frac{\nu(\epsilon_+ -\epsilon_-)}{3w_+(T)}\,, \\
\theta(T) &=& e(T)-\frac{p(T)}{c_s^2}\,.
\eea
$\alpha_\theta$ can also be expressed in terms of $\alpha_p$ and $\psi$ defined in Eqs.\ (\ref{eq:alphaP}) and (\ref{eq:psi}) as $\alpha_\theta(T)=\alpha_p(T)+(1-\psi)/3$. It satisfies the relation
\be\label{eq:alphaTempChange}
\alpha_\theta(T_1)=\alpha_\theta(T_2)\frac{w_+(T_2)}{w_+(T_1)}\,.
\ee

For subsonic solutions, the wall is a deflagration which features a shock wave propagating in front of it, while the plasma behind it is at rest. This implies $v_-=v_w$. For small $v_w$, the solution of Eq.\ (\ref{eq:matching}) is 
\be\label{eq:vPlus}
v_+ \approx (1-3\alpha_\theta(T_+))v_w\,.
\ee

We now wish to express $\alpha_\theta(T_+)$ in terms of $\alpha_\theta(T_*)$, which requires the ratio $w_+(T_+)/w_+(T_*)$. To compute it, one must integrate the conservation of energy-momentum across the shock wave. Away from the wall, the fluid profile is a function of the variable $\xi=r/t$, with $r$ the bubble radius and $t$ the time since nucleation. Notice that in these coordinates, the wall is located at $\xi_w=v_w$. The fluid velocity and enthalpy then satisfy \cite{Espinosa:2010hh}
\bea
2\frac{v}{\xi}&=&\gamma^2(1-v\xi)\left[\frac{1}{c_s^2}\left(\frac{\xi-v}{1-v\xi}\right)^2-1\right]\partial_\xi v\,,\\
\partial_\xi w&=&\nu w\gamma^2\left(\frac{\xi-v}{1-v\xi}\right)\partial_\xi v\,,
\eea
where $v(\xi)$ is now expressed in the frame of the bubble's center. At small $v_w$, these equations can be linearized in terms of $v(\xi)$, which gives the solution
\bea
\label{eq:vSolution}
v(\xi)&\approx& (v_w-v_+)v_w^2\left(\frac{1}{\xi^2}-\frac{1}{c_s^2}\right),\\
\frac{w(\xi)}{w_+(T_*)}&\approx& 1+\frac{\nu}{2\xi^4}v_w^2(v_w-v_+)[4\xi^3-v_w^2(v_w-v_+)]\,,\nn\\
\label{eq:wSolution}
\eea
where we enforced the boundary conditions $v(\xi=v_w)=(v_w-v_+)/(1-v_w v_+)+\mathcal{O}(v_w^3)$ and $w(\xi=c_s)=w_+(T_*)+\mathcal{O}(v_w^3)$.
Finally, the Eqs.\ (\ref{eq:alphaTempChange}), (\ref{eq:vPlus}) and (\ref{eq:wSolution}) imply
\bea
\frac{w_+(T_+)}{w_+(T_*)}&=&\frac{w(\xi=v_w)}{w_+(T_*)}\\
&\approx&1-\frac{3}{2}\nu v_w^2\alpha_\theta(T_*)(3\alpha_\theta(T_*)-4)\nn
\eea
and
\bea\label{eq:alphaTplus}
\alpha_\theta(T_+)&=&\alpha_\theta(T_*)\frac{w_+(T_*)}{w_+(T_+)}\\
&\approx& \alpha_\theta(T_*) \left[1+\frac{3}{2}\nu v_w^2\alpha_\theta(T_*)(3\alpha_\theta(T_*)-4)\right]\,.\nn
\eea\\
This will be useful in what follows.

\paragraph*{\textbf{Kinetic energy fraction.}}
An important parameter needed to compute the GW spectrum is the kinetic energy fraction defined as \cite{Caprini:2019egz}
\be
K=\frac{3}{e_+(T_*)v_w^3}\int_{v_w}^{c_s}\! d\xi\, \xi^2\gamma^2v^2 w\,.
\ee
Here we use the convention where the vacuum energy vanishes inside the bubble, such that $\epsilon_-=0$.
Inserting the solutions (\ref{eq:vPlus}), (\ref{eq:vSolution}) and (\ref{eq:wSolution}) and expanding to lowest order in $v_w$, one finds
\be
K\approx 27v_w^2\alpha_\theta^2(T_*)\frac{w_+(T_*)}{e_+(T_*)}\,.
\ee
Notice that in the LTE and ballistic approximations derived below, $\alpha_p(T_*)$ is at least of order $v_w$, such that $\alpha_\theta(T_*)=(1-\psi)/3+\mathcal{O}(v_w)$ and $p_+(T_*)= p_-(T_*)+\mathcal{O}(v_w)$. Using the relations $w=e+p$ and $w_-=\nu p_-$, one can then show
\be
\frac{e_+}{w_+}\approx 1-\frac{p_-}{w_+}=1-\frac{\psi}{\nu}\,.
\ee
Therefore, the kinetic energy fraction can be simplified to
\be
K\approx \frac{3\nu(1-\psi)^2 v_w^2}{\nu-\psi}\,.
\ee

\paragraph*{\textbf{LTE wall speed.}}

In local thermal equilibrium, the wall velocity can be determined by requiring conservation of entropy. This imposes a new matching condition that must be satisfied across the wall:
\be
s_- \gamma_- v_- = s_+ \gamma_+ v_+\,,
\ee
with $s_\pm=dp_\pm/dT$ the entropy density. Using the template model (\ref{eq:templateModel}), Ref.\ \cite{Ai:2023see} showed that this matching equation can be expressed as
\be
\frac{3\nu\alpha_\theta(T_+) v_+ v_-}{1-(\nu-1)v_+ v_-} = 1-3\alpha_\theta(T_+)-\psi\left(\frac{\gamma_+}{\gamma_-}\right)^\nu.
\ee
Again, this equation can be expanded to the lowest order in the plasma velocities and solved for $v_w=v_-$. To do this, one must first substitute the value of $\alpha_\theta(T_+)$ from Eq.\ (\ref{eq:alphaTplus}) and set $\alpha_\theta(T_*)=(1-\psi)/3+\alpha_p(T_*)$. Assuming $\alpha_p(T_*)$ to be proportional to $v_w^2$ (which is confirmed by the solution below), one finally finds
\be
v_w^{\rm LTE} \approx \frac{c_s}{1-\psi}\sqrt{\frac{2\alpha_p(T_*)}{1+c_s^2}}\,.
\ee

\paragraph*{\textbf{Ballistic wall speed.}}

To obtain a lower bound on the wall velocity, one can consider the situation where a fraction $1-\psi$ of the plasma is completely reflected in front of the wall. The momentum transfer between the wall and the plasma is thus maximized, which leads to a maximal pressure and minimal $v_w$. If the plasma does not equilibrate quickly, this can be described by the large-mass ballistic limit. It was shown in Ref.\ \cite{Ai:2024btx} that this limit yields the condition
\be
\alpha_\theta(T_+)=\frac{1-\psi}{3}(1+v_+)^3\gamma_+^2\,.
\ee
As before, we substitute Eqs.\ (\ref{eq:vPlus}) and (\ref{eq:alphaTplus}) and expand for small $v_w$. Considering that this limit gives a relation $\alpha_p(T_*)\propto v_w$, the solution is
\be
v_w^{\rm ball} \approx \frac{\alpha_p(T_*)}{\psi(1-\psi)}\,,
\ee
as we approximated in Eq.\ (\ref{vwballeq}).

\section{Realistic QCD equation of state}
\label{app:qcdeos}

\begin{figure}
    \centering
    \includegraphics[width=1\linewidth]{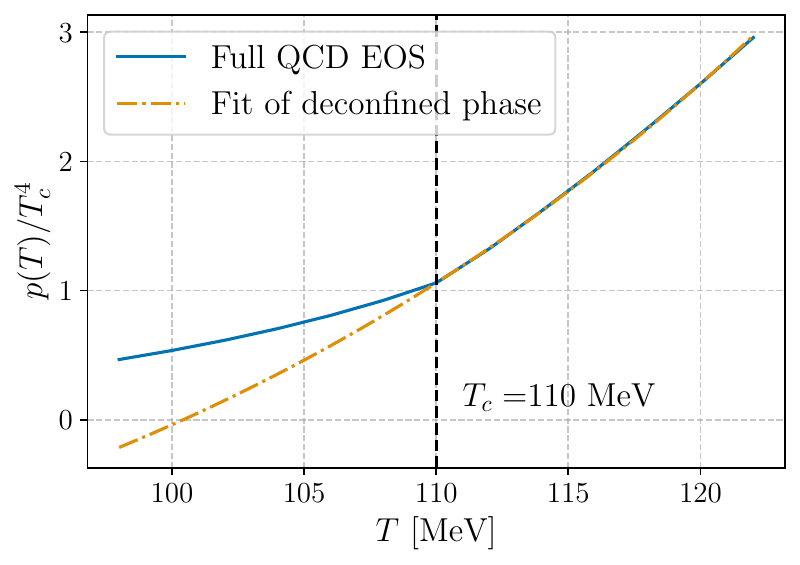}
    \caption{Numerical fit and extrapolation of the deconfined phase for $T_c=110$ MeV. The solid blue line shows the pressure of the most stable phase obtained from the data of Ref.\ \cite{Lu:2023msn}. The dashed yellow line shows the numerical fit of the deconfined phase at $T>T_c$ and its extrapolation to the supercooled regime $T<T_c$}
    \label{fig:highTFit}
\end{figure}

\begin{figure}
    \centering
    \includegraphics[width=1\linewidth]{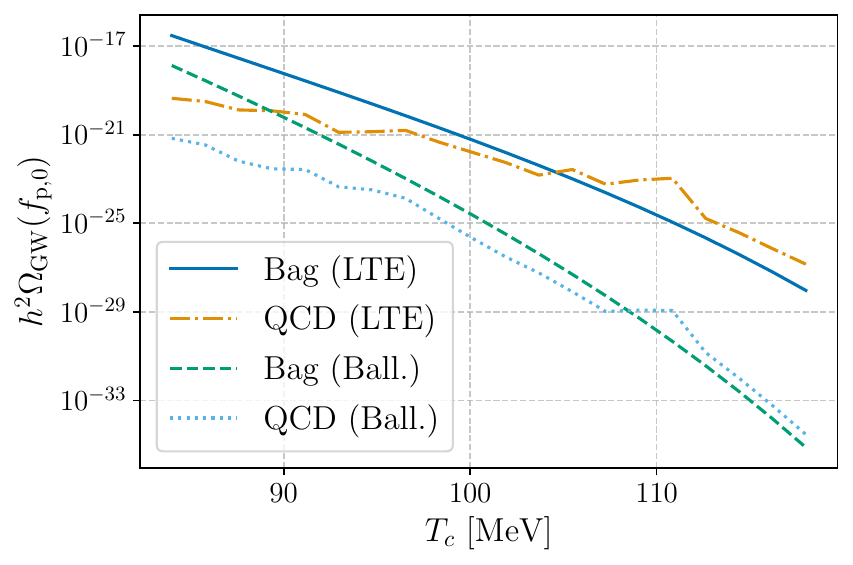}
    \caption{GW amplitude computed using the bag EOS or the full QCD EOS and estimating the wall velocity with the LTE or ballistic approximation.}
    \label{fig:GWComparison}
\end{figure}

Most of the results presented in this work depend on the state of the plasma around the bubble wall, which can be described mathematically by an EOS. To simplify the calculations, we have approximated the full QCD EOS in both phases by the simpler template model presented in Appendix \ref{app:hydro}. It is then natural to ask if our results are sensitive to this approximation and if our conclusions would still hold if a more realistic EOS were used. 

To answer this question, we can repeat our calculations using the data of Ref.\ \cite{Lu:2023msn}, who computed the full QCD EOS at finite temperature and density using a combination of lattice and functional QCD approaches.

One difficulty of working with such a dataset is that it only shows the state of the most stable phase, which corresponds to the deconfined and confined phases at $T>T_c$ and $T<T_c$, respectively. However, since we expect some degree of supercooling, we also need a description the deconfined phase when $T<T_c$. To obtain it, one can fit the template model (\ref{eq:templateModel}) to the EOS data of Ref.\ \cite{Lu:2023msn} when $T>T_c$, and then extrapolate this fit to the supercooled regime $T<T_c$. An example of such a fit is shown in Fig.\ \ref{fig:highTFit}.

Using this extrapolation to describe the supercooled deconfined phase and the EOS data of Ref.\ \cite{Lu:2023msn} to describe the confined phase, it is now possible to compute the GW spectrum with a straightforward generalization of the method presented in this work\footnote{The wall velocities in the LTE and ballistic approximations were estimated by computing the sound speeds in both phases and approximating the EOS by the template model.}. A comparison of the GW amplitude computed with different approximations is shown in Fig.\ \ref{fig:GWComparison}. One can see that at high $T_c$, the bag EOS yields a good approximation to the more realistic QCD EOS. However, at lower $T_c$, the larger supercooling makes the bag EOS less accurate and leads to an overestimation of the GW amplitude. One can thus see the bag EOS as an optimistic approximation, which confirms that the GWs produced in that temperature range would not be seen by $\mu$Ares.

\end{appendix}

\bibliography{references}

\providecommand{\href}[2]{#2}\begingroup\raggedright\begin{thebibliography}{10}

\bibitem{Laurent:2022jrs}
B.~Laurent and J.~M. Cline, ``{First principles determination of bubble wall velocity},'' \href{http://dx.doi.org/10.1103/PhysRevD.106.023501}{{\em Phys. Rev. D} {\bfseries 106} no.~2, (2022) 023501}, \href{http://arxiv.org/abs/2204.13120}{{\ttfamily arXiv:2204.13120 [hep-ph]}}.

\bibitem{Ekstedt:2024fyq}
A.~Ekstedt, O.~Gould, J.~Hirvonen, B.~Laurent, L.~Niemi, P.~Schicho, and J.~van~de Vis, ``{How fast does the WallGo? A package for computing wall velocities in first-order phase transitions},'' \href{http://arxiv.org/abs/2411.04970}{{\ttfamily arXiv:2411.04970 [hep-ph]}}.

\bibitem{Dorsch:2024jjl}
G.~C. Dorsch, T.~Konstandin, E.~Perboni, and D.~A. Pinto, ``{Non-singular solutions to the Boltzmann equation with a fluid Ansatz},'' \href{http://arxiv.org/abs/2412.09266}{{\ttfamily arXiv:2412.09266 [hep-ph]}}.

\bibitem{Ai:2021kak}
W.-Y. Ai, B.~Garbrecht, and C.~Tamarit, ``{Bubble wall velocities in local equilibrium},'' \href{http://dx.doi.org/10.1088/1475-7516/2022/03/015}{{\em JCAP} {\bfseries 03} no.~03, (2022) 015}, \href{http://arxiv.org/abs/2109.13710}{{\ttfamily arXiv:2109.13710 [hep-ph]}}.

\bibitem{Ai:2023see}
W.-Y. Ai, B.~Laurent, and J.~van~de Vis, ``{Model-independent bubble wall velocities in local thermal equilibrium},'' \href{http://dx.doi.org/10.1088/1475-7516/2023/07/002}{{\em JCAP} {\bfseries 07} (2023) 002}, \href{http://arxiv.org/abs/2303.10171}{{\ttfamily arXiv:2303.10171 [astro-ph.CO]}}.

\bibitem{Sanchez-Garitaonandia:2023zqz}
M.~Sanchez-Garitaonandia and J.~van~de Vis, ``{Prediction of the bubble wall velocity for a large jump in degrees of freedom},'' \href{http://dx.doi.org/10.1103/PhysRevD.110.023509}{{\em Phys. Rev. D} {\bfseries 110} no.~2, (2024) 023509}, \href{http://arxiv.org/abs/2312.09964}{{\ttfamily arXiv:2312.09964 [hep-ph]}}.

\bibitem{Ai:2024btx}
W.-Y. Ai, B.~Laurent, and J.~van~de Vis, ``{Bounds on the bubble wall velocity},'' \href{http://dx.doi.org/10.1007/JHEP02(2025)119}{{\em JHEP} {\bfseries 02} (2025) 119}, \href{http://arxiv.org/abs/2411.13641}{{\ttfamily arXiv:2411.13641 [hep-ph]}}.

\bibitem{Krajewski:2024zxg}
T.~Krajewski, M.~Lewicki, I.~Nalecz, and M.~Zych, ``{Steady-state bubbles beyond local thermal equilibrium},'' \href{http://arxiv.org/abs/2411.16580}{{\ttfamily arXiv:2411.16580 [astro-ph.CO]}}.

\bibitem{Schwarz:2009ii}
D.~J. Schwarz and M.~Stuke, ``{Lepton asymmetry and the cosmic QCD transition},'' \href{http://dx.doi.org/10.1088/1475-7516/2009/11/025}{{\em JCAP} {\bfseries 11} (2009) 025}, \href{http://arxiv.org/abs/0906.3434}{{\ttfamily arXiv:0906.3434 [hep-ph]}}. [Erratum: JCAP 10, E01 (2010)].

\bibitem{Wygas:2018otj}
M.~M. Wygas, I.~M. Oldengott, D.~B\"odeker, and D.~J. Schwarz, ``{Cosmic QCD Epoch at Nonvanishing Lepton Asymmetry},'' \href{http://dx.doi.org/10.1103/PhysRevLett.121.201302}{{\em Phys. Rev. Lett.} {\bfseries 121} no.~20, (2018) 201302}, \href{http://arxiv.org/abs/1807.10815}{{\ttfamily arXiv:1807.10815 [hep-ph]}}.

\bibitem{Gao:2021nwz}
F.~Gao and I.~M. Oldengott, ``{Cosmology Meets Functional QCD: First-Order Cosmic QCD Transition Induced by Large Lepton Asymmetries},'' \href{http://dx.doi.org/10.1103/PhysRevLett.128.131301}{{\em Phys. Rev. Lett.} {\bfseries 128} no.~13, (2022) 131301}, \href{http://arxiv.org/abs/2106.11991}{{\ttfamily arXiv:2106.11991 [hep-ph]}}.

\bibitem{Gao:2024fhm}
F.~Gao, J.~Harz, C.~Hati, Y.~Lu, I.~M. Oldengott, and G.~White, ``{Baryogenesis and first-order QCD transition with gravitational waves from a large lepton asymmetry},'' \href{http://arxiv.org/abs/2407.17549}{{\ttfamily arXiv:2407.17549 [hep-ph]}}.

\bibitem{Gao:2023djs}
F.~Gao, J.~Harz, C.~Hati, Y.~Lu, I.~M. Oldengott, and G.~White, ``{Sphaleron freeze-in baryogenesis with gravitational waves from the QCD transition},'' \href{http://arxiv.org/abs/2309.00672}{{\ttfamily arXiv:2309.00672 [hep-ph]}}.

\bibitem{Ipek:2018lhm}
S.~Ipek and T.~M.~P. Tait, ``{Early Cosmological Period of QCD Confinement},'' \href{http://dx.doi.org/10.1103/PhysRevLett.122.112001}{{\em Phys. Rev. Lett.} {\bfseries 122} no.~11, (2019) 112001}, \href{http://arxiv.org/abs/1811.00559}{{\ttfamily arXiv:1811.00559 [hep-ph]}}.

\bibitem{Gouttenoire:2023roe}
Y.~Gouttenoire, E.~Kuflik, and D.~Liu, ``{Heavy baryon dark matter from SU(N) confinement: Bubble wall velocity and boundary effects},'' \href{http://dx.doi.org/10.1103/PhysRevD.109.035002}{{\em Phys. Rev. D} {\bfseries 109} no.~3, (2024) 035002}, \href{http://arxiv.org/abs/2311.00029}{{\ttfamily arXiv:2311.00029 [hep-ph]}}.

\bibitem{Wu:2006su}
L.-K. Wu, X.-Q. Luo, and H.-S. Chen, ``{Phase structure of lattice QCD with two flavors of Wilson quarks at finite temperature and chemical potential},'' \href{http://dx.doi.org/10.1103/PhysRevD.76.034505}{{\em Phys. Rev. D} {\bfseries 76} (2007) 034505}, \href{http://arxiv.org/abs/hep-lat/0611035}{{\ttfamily arXiv:hep-lat/0611035}}.

\bibitem{JLQCD:1998mja}
{\bfseries JLQCD} Collaboration, S.~Aoki {\em et~al.}, ``{Phase structure of lattice QCD at finite temperature for (2+1) flavors of Kogut-Susskind quarks},'' \href{http://dx.doi.org/10.1016/S0920-5632(99)85104-4}{{\em Nucl. Phys. B Proc. Suppl.} {\bfseries 73} (1999) 459--461}, \href{http://arxiv.org/abs/hep-lat/9809102}{{\ttfamily arXiv:hep-lat/9809102}}.

\bibitem{Guenther:2020jwe}
J.~N. Guenther, ``{Overview of the QCD phase diagram: Recent progress from the lattice},'' \href{http://dx.doi.org/10.1140/epja/s10050-021-00354-6}{{\em Eur. Phys. J. A} {\bfseries 57} no.~4, (2021) 136}, \href{http://arxiv.org/abs/2010.15503}{{\ttfamily arXiv:2010.15503 [hep-lat]}}.

\bibitem{Lu:2023msn}
Y.~Lu, F.~Gao, B.~Fu, H.~Song, and Y.-X. Liu, ``{Constructing the equation of state of QCD in a functional QCD based scheme},'' \href{http://dx.doi.org/10.1103/PhysRevD.109.114031}{{\em Phys. Rev. D} {\bfseries 109} no.~11, (2024) 114031}, \href{http://arxiv.org/abs/2310.16345}{{\ttfamily arXiv:2310.16345 [hep-ph]}}.

\bibitem{Shao:2024dxt}
J.~Shao, H.~Mao, and M.~Huang, ``{The Transition Rate and Gravitational Wave Spectrum from First-Order QCD Phase Transitions},'' \href{http://arxiv.org/abs/2410.06780}{{\ttfamily arXiv:2410.06780 [hep-ph]}}.

\bibitem{Zheng:2024tib}
H.-w. Zheng, F.~Gao, L.~Bian, S.-x. Qin, and Y.-x. Liu, ``{Quantitative analysis of the gravitational wave spectrum sourced from a first-order chiral phase transition of QCD},'' \href{http://dx.doi.org/10.1103/PhysRevD.111.L021303}{{\em Phys. Rev. D} {\bfseries 111} no.~2, (2025) L021303}, \href{http://arxiv.org/abs/2407.03795}{{\ttfamily arXiv:2407.03795 [hep-ph]}}.

\bibitem{Han:2023znh}
X.~Han and G.~Shao, ``{Stochastic gravitational waves produced by the first-order QCD phase transition},'' \href{http://arxiv.org/abs/2312.00571}{{\ttfamily arXiv:2312.00571 [astro-ph.CO]}}.

\bibitem{Feng:2022fwf}
Q.-M. Feng, Z.-W. Feng, X.~Zhou, and Q.-Q. Jiang, ``{Barrow entropy and stochastic gravitational wave background generated from cosmological QCD phase transition},'' \href{http://dx.doi.org/10.1016/j.physletb.2023.137739}{{\em Phys. Lett. B} {\bfseries 838} (2023) 137739}, \href{http://arxiv.org/abs/2210.10658}{{\ttfamily arXiv:2210.10658 [gr-qc]}}.

\bibitem{Zhu:2021vkj}
Z.-R. Zhu, J.~Chen, and D.~Hou, ``{Gravitational waves from holographic QCD phase transition with gluon condensate},'' \href{http://dx.doi.org/10.1140/epja/s10050-022-00754-2}{{\em Eur. Phys. J. A} {\bfseries 58} no.~6, (2022) 104}, \href{http://arxiv.org/abs/2109.09933}{{\ttfamily arXiv:2109.09933 [hep-ph]}}.

\bibitem{Rezapour:2020mvi}
S.~Rezapour, K.~Bitaghsir~Fadafan, and M.~Ahmadvand, ``{Gravitational waves of a first-order QCD phase transition at finite coupling from holography},'' \href{http://dx.doi.org/10.1016/j.aop.2021.168731}{{\em Annals Phys.} {\bfseries 437} (2022) 168731}, \href{http://arxiv.org/abs/2006.04265}{{\ttfamily arXiv:2006.04265 [hep-th]}}.

\bibitem{Caprini:2010xv}
C.~Caprini, R.~Durrer, and X.~Siemens, ``{Detection of gravitational waves from the QCD phase transition with pulsar timing arrays},'' \href{http://dx.doi.org/10.1103/PhysRevD.82.063511}{{\em Phys. Rev. D} {\bfseries 82} (2010) 063511}, \href{http://arxiv.org/abs/1007.1218}{{\ttfamily arXiv:1007.1218 [astro-ph.CO]}}.

\bibitem{Ahmadvand:2017xrw}
M.~Ahmadvand and K.~Bitaghsir~Fadafan, ``{Gravitational waves generated from the cosmological QCD phase transition within AdS/QCD},'' \href{http://dx.doi.org/10.1016/j.physletb.2017.07.039}{{\em Phys. Lett. B} {\bfseries 772} (2017) 747--751}, \href{http://arxiv.org/abs/1703.02801}{{\ttfamily arXiv:1703.02801 [hep-th]}}.

\bibitem{Reichert:2021cvs}
M.~Reichert, F.~Sannino, Z.-W. Wang, and C.~Zhang, ``{Dark confinement and chiral phase transitions: gravitational waves vs matter representations},'' \href{http://dx.doi.org/10.1007/JHEP01(2022)003}{{\em JHEP} {\bfseries 01} (2022) 003}, \href{http://arxiv.org/abs/2109.11552}{{\ttfamily arXiv:2109.11552 [hep-ph]}}.

\bibitem{Huang:2020crf}
W.-C. Huang, M.~Reichert, F.~Sannino, and Z.-W. Wang, ``{Testing the dark SU(N) Yang-Mills theory confined landscape: From the lattice to gravitational waves},'' \href{http://dx.doi.org/10.1103/PhysRevD.104.035005}{{\em Phys. Rev. D} {\bfseries 104} no.~3, (2021) 035005}, \href{http://arxiv.org/abs/2012.11614}{{\ttfamily arXiv:2012.11614 [hep-ph]}}.

\bibitem{Shao:2022oqw}
J.~Shao and M.~Huang, ``{Gravitational waves and primordial black holes from chirality imbalanced QCD first-order phase transition with P and CP violation},'' \href{http://dx.doi.org/10.1103/PhysRevD.107.043011}{{\em Phys. Rev. D} {\bfseries 107} no.~4, (2023) 043011}, \href{http://arxiv.org/abs/2209.13809}{{\ttfamily arXiv:2209.13809 [hep-ph]}}.

\bibitem{Li:2018oqf}
M.-W. Li, Y.~Yang, and P.-H. Yuan, ``{Imprints of Early Universe on Gravitational Waves from First-Order Phase Transition in QCD},'' \href{http://arxiv.org/abs/1812.09676}{{\ttfamily arXiv:1812.09676 [hep-th]}}.

\bibitem{Ahmadvand:2017tue}
M.~Ahmadvand and K.~Bitaghsir~Fadafan, ``{The cosmic QCD phase transition with dense matter and its gravitational waves from holography},'' \href{http://dx.doi.org/10.1016/j.physletb.2018.01.066}{{\em Phys. Lett. B} {\bfseries 779} (2018) 1--8}, \href{http://arxiv.org/abs/1707.05068}{{\ttfamily arXiv:1707.05068 [hep-th]}}.

\bibitem{Huber:2007vva}
S.~J. Huber and T.~Konstandin, ``{Production of gravitational waves in the nMSSM},'' \href{http://dx.doi.org/10.1088/1475-7516/2008/05/017}{{\em JCAP} {\bfseries 05} (2008) 017}, \href{http://arxiv.org/abs/0709.2091}{{\ttfamily arXiv:0709.2091 [hep-ph]}}.

\bibitem{Leitao:2014pda}
L.~Leitao and A.~Megevand, ``{Hydrodynamics of phase transition fronts and the speed of sound in the plasma},'' \href{http://dx.doi.org/10.1016/j.nuclphysb.2014.12.008}{{\em Nucl. Phys. B} {\bfseries 891} (2015) 159--199}, \href{http://arxiv.org/abs/1410.3875}{{\ttfamily arXiv:1410.3875 [hep-ph]}}.

\bibitem{PhysRevC.100.024907}
A.~Monnai, B.~Schenke, and C.~Shen, ``Equation of state at finite densities for qcd matter in nuclear collisions,'' \href{http://dx.doi.org/10.1103/PhysRevC.100.024907}{{\em Phys. Rev. C} {\bfseries 100} (Aug, 2019) 024907}. \url{https://link.aps.org/doi/10.1103/PhysRevC.100.024907}.

\bibitem{Bresciani:2025vxw}
M.~Bresciani, M.~D. Brida, L.~Giusti, and M.~Pepe, ``{The QCD Equation of State with $N_f=3$ flavours up to the electro-weak scale},'' \href{http://arxiv.org/abs/2501.11603}{{\ttfamily arXiv:2501.11603 [hep-lat]}}.

\bibitem{Gao:2016hks}
F.~Gao and Y.-x. Liu, ``{Interface Effect in QCD Phase Transitions via Dyson-Schwinger Equation Approach},'' \href{http://dx.doi.org/10.1103/PhysRevD.94.094030}{{\em Phys. Rev. D} {\bfseries 94} no.~9, (2016) 094030}, \href{http://arxiv.org/abs/1609.08038}{{\ttfamily arXiv:1609.08038 [hep-ph]}}.

\bibitem{Guo:2020grp}
H.-K. Guo, K.~Sinha, D.~Vagie, and G.~White, ``{Phase Transitions in an Expanding Universe: Stochastic Gravitational Waves in Standard and Non-Standard Histories},'' \href{http://dx.doi.org/10.1088/1475-7516/2021/01/001}{{\em JCAP} {\bfseries 01} (2021) 001}, \href{http://arxiv.org/abs/2007.08537}{{\ttfamily arXiv:2007.08537 [hep-ph]}}.

\bibitem{Hindmarsh:2020hop}
M.~B. Hindmarsh, M.~L\"uben, J.~Lumma, and M.~Pauly, ``{Phase transitions in the early universe},'' \href{http://dx.doi.org/10.21468/SciPostPhysLectNotes.24}{{\em SciPost Phys. Lect. Notes} {\bfseries 24} (2021) 1}, \href{http://arxiv.org/abs/2008.09136}{{\ttfamily arXiv:2008.09136 [astro-ph.CO]}}.

\bibitem{Sesana:2019vho}
A.~Sesana {\em et~al.}, ``{Unveiling the gravitational universe at $\mu$-Hz frequencies},'' \href{http://dx.doi.org/10.1007/s10686-021-09709-9}{{\em Exper. Astron.} {\bfseries 51} no.~3, (2021) 1333--1383}, \href{http://arxiv.org/abs/1908.11391}{{\ttfamily arXiv:1908.11391 [astro-ph.IM]}}.

\bibitem{Bea:2021zsu}
Y.~Bea, J.~Casalderrey-Solana, T.~Giannakopoulos, D.~Mateos, M.~Sanchez-Garitaonandia, and M.~Zilh\~ao, ``{Bubble wall velocity from holography},'' \href{http://dx.doi.org/10.1103/PhysRevD.104.L121903}{{\em Phys. Rev. D} {\bfseries 104} no.~12, (2021) L121903}, \href{http://arxiv.org/abs/2104.05708}{{\ttfamily arXiv:2104.05708 [hep-th]}}.

\bibitem{Bigazzi:2021ucw}
F.~Bigazzi, A.~Caddeo, T.~Canneti, and A.~L. Cotrone, ``{Bubble wall velocity at strong coupling},'' \href{http://dx.doi.org/10.1007/JHEP08(2021)090}{{\em JHEP} {\bfseries 08} (2021) 090}, \href{http://arxiv.org/abs/2104.12817}{{\ttfamily arXiv:2104.12817 [hep-ph]}}.

\bibitem{Giese:2020rtr}
F.~Giese, T.~Konstandin, and J.~van~de Vis, ``{Model-independent energy budget of cosmological first-order phase transitions\textemdash{}A sound argument to go beyond the bag model},'' \href{http://dx.doi.org/10.1088/1475-7516/2020/07/057}{{\em JCAP} {\bfseries 07} no.~07, (2020) 057}, \href{http://arxiv.org/abs/2004.06995}{{\ttfamily arXiv:2004.06995 [astro-ph.CO]}}.

\bibitem{Espinosa:2010hh}
J.~R. Espinosa, T.~Konstandin, J.~M. No, and G.~Servant, ``{Energy Budget of Cosmological First-order Phase Transitions},'' \href{http://dx.doi.org/10.1088/1475-7516/2010/06/028}{{\em JCAP} {\bfseries 06} (2010) 028}, \href{http://arxiv.org/abs/1004.4187}{{\ttfamily arXiv:1004.4187 [hep-ph]}}.

\bibitem{Caprini:2019egz}
C.~Caprini {\em et~al.}, ``{Detecting gravitational waves from cosmological phase transitions with LISA: an update},'' \href{http://dx.doi.org/10.1088/1475-7516/2020/03/024}{{\em JCAP} {\bfseries 03} (2020) 024}, \href{http://arxiv.org/abs/1910.13125}{{\ttfamily arXiv:1910.13125 [astro-ph.CO]}}.

\end{thebibliography}\endgroup
\bibliographystyle{utphys}

\end{document}